\documentclass[conference]{IEEEtran}
\IEEEoverridecommandlockouts
\usepackage{cite}
\usepackage{amsmath,amssymb,amsfonts}
\usepackage[T1]{fontenc}
\usepackage{algorithmic}
\usepackage{graphicx}
\usepackage{textcomp}
\usepackage{xcolor}
\usepackage{arydshln}
\usepackage{multirow}
\usepackage{cite}
\usepackage[normalem]{ulem}
\usepackage{float}
\def\BibTeX{{\rm B\kern-.05em{\sc i\kern-.025em b}\kern-.08em
    T\kern-.1667em\lower.7ex\hbox{E}\kern-.125emX}}

\usepackage{ulem}

\begin{document}

\title{HIGH-RESOLUTION SPEAKER COUNTING IN REVERBERANT ROOMS USING CRNN WITH AMBISONICS FEATURES\\
}

\author{\IEEEauthorblockN{Pierre-Amaury Grumiaux}
\IEEEauthorblockA{\textit{Orange Labs}\\
Cesson-S{\'e}vign{\'e}, France \\
pierreamaury.grumiaux@orange.com}
\and
\IEEEauthorblockN{Sr\dj{}an Kiti\'c}
\IEEEauthorblockA{
\textit{Orange Labs}\\
Cesson-S{\'e}vign{\'e}, France \\
srdan.kitic@orange.com}
\and
\IEEEauthorblockN{Laurent Girin}
\IEEEauthorblockA{
\textit{Univ. Grenoble Alpes, GIPSA-lab}\\
\textit{Grenoble-INP, CNRS}\\
Grenoble, France \\
laurent.girin@grenoble-inp.fr}
\and
\IEEEauthorblockN{Alexandre Gu{\'e}rin}
\IEEEauthorblockA{
\textit{Orange Labs}\\
Cesson-S{\'e}vign{\'e}, France \\
alexandre.guerin@orange.com}
}

\maketitle

\begin{abstract}
Speaker counting is the task of estimating the number of people that are simultaneously speaking in an audio recording. For several audio processing tasks such as speaker diarization, separation, localization and tracking, knowing the number of speakers at each timestep is a prerequisite, or at least it can be a strong advantage, in addition to enabling a low latency processing. For that purpose, we address the speaker counting problem with a multichannel convolutional recurrent neural network which produces an estimation at a short-term frame resolution. We trained the network to predict up to $5$ concurrent speakers in a multichannel mixture, with simulated data including many different conditions in terms of source and microphone positions, reverberation, and noise. The network can predict the number of speakers with good accuracy at frame resolution.
% SK: fair == decent, thus too weak for an abstract.
\end{abstract}

\begin{IEEEkeywords}
Speaker counting, CRNN, reverberation.
\end{IEEEkeywords}

\section{Introduction}

Speaker counting is the task of estimating the number of people that are simultaneously speaking in successive segments of an audio recording. It can be seen as a subtask of speaker diarization, which consists in estimating who speaks and when, and which has long been limited to the case where one person speaks at a time, since it becomes much more complicated when several speech signals overlap \cite{Tranteroverviewautomaticspeaker2006,AngueraSpeakerDiarizationReview2012}.

Although speaker counting has been relatively poorly addressed in the speech processing literature as a problem on its own, it can be an essential primary module for more complex machine audition tasks, in particular for source separation, localisation and tracking. 
Yet, the vast majority of speech/audio source separation and localisation methods either consider that the number of sources to process is known a priori or has been previously estimated \cite{VincentProbabilisticmodelingparadigms2011,MakinoBlindSpeechSeparation2007,ValinRobustLocalizationTracking2007,PerotinCRNNBasedMultipleDoA2019}, or it is estimated from some clustering of the separation/localisation features, or they consider a maximum number of speakers \cite{FallonAcousticSourceLocalization2012a,LiOnlineLocalizationTracking2019}.
Speaker counting is also particularly useful for tracking, as it can help solving the difficult problem of detecting the appearance and disappearance of a speaker track along time as he/she starts/stops speaking \cite{VoMultitargetTracking2015}. 

In the literature, a few single-channel parametric methods for speaker counting correlate the number of speakers to some ad-hoc features extracted from the mixture signal {\cite{SayoudProposalNewConfidence2010},\cite{AraiEstimatingnumberspeakers2003}. These methods fail at exploiting the spatial discrimination between the different sound sources. Classical multichannel approaches, based on the eigenvalue analysis of the estimated spatial covariance matrix (SCM), take this information into account \cite{WilliamsDetectionDeterminingNumber1999, NadakuditiSampleEigenvalueBased2008, YamamotoEstimationnumbersound2003}. However, they cannot be used in an underdetermined setup.
Methods based on clustering in the time-frequency (TF) domain remove this restriction  \cite{balan2007estimator,ArberetRobustMethodCount2010,HiguchiUnifiedapproachaudio2015, Kounades-BastianEMalgorithmjoint2017, YangMultipleSoundSource2019}, but are generally limited to the anechoic setting, and often require additional a priori information, \emph{e.g.} the maximum number of concurrent sources in the processed sequence.

More recently, some attempts have been made to apply deep learning to audio source counting. In \cite{WeiDeterminingNumberSpeakers2018}, a convolutional neural network is used to classify single-channel noisy audio signals into 3 classes: 1, 2, and 3-or-more sources. In \cite{StoterClassificationvsRegression2018}, St{\"o}ter et al. compared several representations in supervised learning for single-channel speaker counting. The best results were obtained with a bi-directional long short-term memory (bi-LSTM) neural network, short-time Fourier transform (STFT) features, and a classification configuration. This work was extended in \cite{StoterCountNetEstimatingNumber2019} with a convolutional recurrent neural network (CRNN) named \textit{CountNet}, which has shown superior performance against traditional methods for estimating the maximum number of simultaneous speakers within an audio excerpt of 5 seconds. The importance of learning from reverberant audio examples was discussed, as well as the extension to a multichannel setup, although the latter has not been investigated. Therefore, being essentially a single-channel method, CountNet is blind to the spatial aspect of the source counting problem.

The contributions of the present paper are twofold:
\begin{itemize}
    \item First, we evaluate the benefit of using a multichannel input in a neural network for the speaker counting problem. In the present study, we use the Ambisonics multichannel audio format, due to its increasing interest in interactive spatial audio applications, like Facebook 360\footnote{https://facebookincubator.github.io/facebook-360-spatial-workstation/Documentation/SpatialWorkstation/SpatialWorkstation.html, visited on 02/03/2020} or YouTube\footnote{https://support.google.com/youtube/answer/6395969, visited on 02/03/2020}
    \item Second, we tackle the challenging problem of estimating the number of speakers at a short-term frame resolution. Compared to usual estimation on longer segments, this is a crucial novelty and advantage for exploiting speaker counting in further processes such as speaker separation and localization, since those latter are generally processed on a short-term frame-by-frame basis. Also, this would enable a low-latency (possibly real-time) overall process.
    % I have decided not to list the low-latency (4-frame-delay) property of our method, in order to keep some material for Forum Acousticum
\end{itemize}
We show that the proposed CRNN with Ambisonics multichannel inputs leads to improved framewise counting performance upon a monochannel CRNN, and a state-of-the-art multichannel method. 
\vspace*{-0.2cm}
\section{Proposed method}
\label{sec:method}

We now successively describe the input features, the output configuration, the mapping strategy and the network architecture we used for speaker counting.

\vspace*{-0.1cm}
\subsection{Input features}

In the present study, we use the Ambisonics signal representation as a multichannel input of our neural network. The Ambisonics format} is particularly well-suited to represent the spatial properties of a soundfield, and is, to some extent, agnostic to the microphone array configuration \cite{DanielRepresentationchampsacoustiques2001}. 
That said, we do not claim that this representation is better than other (more conventional) multichannel formats for the speaker counting task, its use is here a choice of convenience in a general and more and more popular applicative framework.

The Ambisonics format is produced by projecting the recorded multichannel audio onto a basis of spherical harmonic functions. The number of retained coefficients defines the Ambisonics order: in practice, to obtain an Ambisonics representation of order $m$, one needs a spherical microphone array containing at least $(m+1)^2$ capsules. Since, in theory, the capsules would need to be perfectly coincident, a set of phase calibration filters is applied beforehand. The use of first-order Ambisonics (FOA) ($m=1$) has been shown to provide a neural network with sufficient spatial information for single- and multi-speaker localization \cite{PerotinCRNNbasedjointazimuth2018,PerotinCRNNBasedMultipleDoA2019}, thus motivating our choice as the input features. FOA provides a decomposition of the signal into the first four Ambisonics channels denoted $W$, $X$, $Y$, $Z$. Channel $W$ (order-0 spherical harmonic) represents the soundfield as if it was recorded by an omnidirectional microphone at the observation point. 
Channels $X$, $Y$ and $Z$ (order-1 spherical harmonics) correspond to the recordings of three polarized orthogonal bidirectional microphones. For a plane wave coming from azimuth $\theta$ and elevation $\phi$, and bearing a sound pressure $p$, the FOA components are given in the STFT domain by:\footnote{We adopt the N3D Ambisonics normalization standard \cite{DanielRepresentationchampsacoustiques2001}.}
\begin{equation}\label{eqFOA}
\begin{bmatrix} W(t,f) \\ X(t,f) \\ Y(t,f) \\ Z(t,f) \end{bmatrix} = \begin{bmatrix} 1 \\ 
\sqrt{3} \cos\theta \cos\phi \\ \sqrt{3} \sin\theta \cos\phi \\ \sqrt{3} \sin\phi \end{bmatrix} p(t,f).
\end{equation}
where $t$ and $f$ denote the STFT time and frequency bins, respectively.

Because the phase of $p(t,f)$ is a common information across channels, it does not provide much information for the spatial discrimination of different speakers. Taking the magnitude of the FOA vector entries only discards the sign of the trigonometric terms, which leads to ambiguities only for specific spatial configurations. In short, the spatial information of the FOA channels is mostly encoded in their magnitude, We thus select this input representation: the magnitude of the STFT of the four FOA channels is computed and stacked to give a tridimensional tensor $\mathbf{X} \in \mathbb{R}^{N_t \times F \times I}$ with $N_t$ time frames, $F$ frequency bins and $I=4$ channels, which is the input feature for the neural network. The role of the number of frames $N_t$, which is part of the parameters tested in our experiments, will be detailed in Section~\ref{seq2seq}. In our experiments, we use signals sampled at $16$~kHz, a $1$,$024$-point STFT (hence $F=513$) with a sinusoidal analysis window and $50\%$ overlap. One  frame thus represents $32$\,ms of additional signal information.

%\vspace*{-0.1cm}
\subsection{Outputs}
\label{outputs}

We consider speaker counting as a classification problem where each class corresponds to a different number of active speakers from $0$ (i.e. only background noise) to a maximum of $5$ active speakers. 
During training, the output target encoding the class probabilities is a one-hot vector $\mathbf{y}$ of size $6$. The softmax function is used at the output layer of the neural network, to represent the probability distribution over the $6$ classes. 
The predicted number of speakers is the class with the highest output probability.
For training we use the categorical cross-entropy loss.

\subsection{Sequence-to-sequence mapping}
\label{seq2seq}

In \cite{StoterCountNetEstimatingNumber2019}, the \emph{maximum} number of active speakers within a $5$-s segment, i.e. a large sequence of short-term frames, was estimated. The whole segment was labeled with this maximum number of speakers, and a unique sequence-to-one decoding scheme was used to estimate this number for each segment. In contrast, in the present work, we target a much finer temporal resolution: We aim at predicting the total number of speakers for each short-term frame.\footnote{Of course, speaker counting with a lower time resolution can then be obtained by filtering the frame-wise results (e.g. with majority voting).} To this aim, for training, each short-term frame is labeled by the total number of active speakers within the frame. Although we consider frame resolution at decoding, we still want to exploit a larger local context, but at a lower scale than the 5\,s of \cite{StoterCountNetEstimatingNumber2019}: For each frame to classify, we use a short signal sequence of $N_t$ frames as corresponding input to the network. $N_t$ is within 10 to 30, i.e. a $320$-ms to $1$-s local context in our experiments. We treat the problem as a sequence-to-sequence scheme, combined with one-frame shift. Each input sequence of $N_t$ frames produces a synchronized sequence of $N_t$ decoded class probability vectors. We actually classify only the last frame within a sequence before proceeding to one-frame shift, and repeating the process. Our experiments demonstrated that the best decoding performance for that last frame was obtained by selecting the output vector at position $N_t - 4$ in the decoded sequence, which we adopt hereafter. Detailed analysis of this process will be reported in \cite{ForumAcusticum}.  

%\LG{}{[]PARAGRAPH CI-DESSOUS A SUPPRIMER OU BIEN A MIXER AVEC LE PRECEDENT]}
%\SK{}{During inference, the audio data is streamed, in which case the network is continuously fed by the input tensors $\mathbf{X}_0, \mathbf{X}_1, \mathbf{X}_2,\hdots\mathbf{X}_j,\hdots$. Here, $j$ denotes the frame index, meaning that the inputs largely overlap. This enables us to select -- among $N_t$ output labels -- the optimal one with respect to predicting the number of speakers in the latest frame $j$ (while maintaining the overall low processing latency). In \cite{ForumAcousticum}, we demonstrate that this corresponds to the label $N_t - 4$, which we adopt hereafter.} 

%\vspace*{-0.1cm}
\subsection{Network architecture}

To design our network, we took inspiration from  the CountNet CRNN architecture \cite{StoterCountNetEstimatingNumber2019} (itself inspired from \cite{leglaive2015singing}), which was shown to be effective in exploiting the spectral information within a single-channel mixture. We speculate that this type of network can benefit from spatial information provided by the multichannel input.

\begin{figure}[t]
    \centering
    \includegraphics[width=0.6\columnwidth]{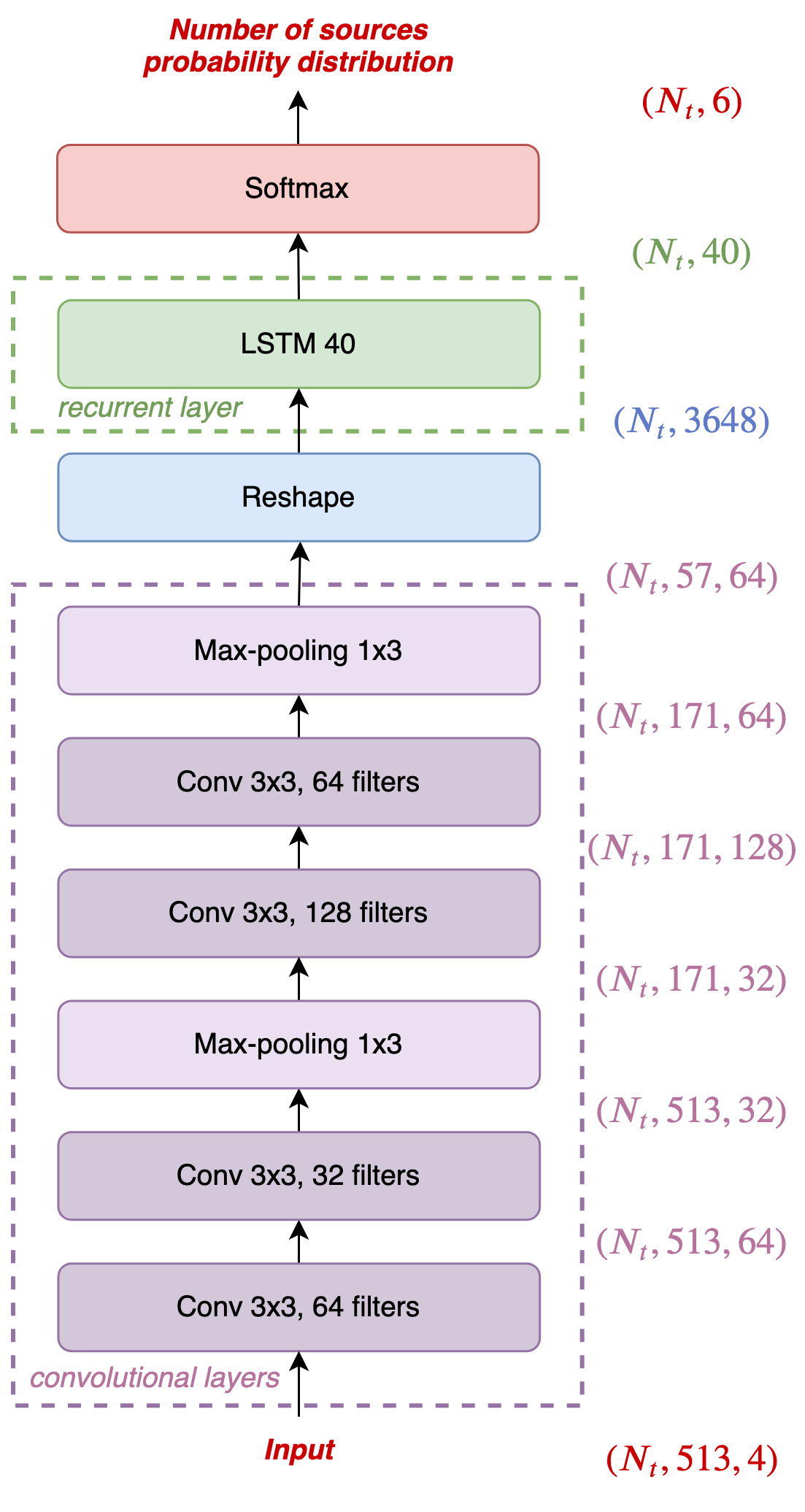}
    \caption{Architecture of the proposed speaker counting neural network. The dimension of each layer's output is also given.}
    \label{fig:crnn}
\end{figure}

The resulting architecture is illustrated in Fig.~\ref{fig:crnn}. The first part of the network is composed of two convolutional layers with $64$ and $32$, $3\!\times\!3$ filters per channel, respectively (applied in the time-frequency dimensions), ending by an $1 \times 3$ max-pooling layer. This is followed by another two convolutional layers with $128$ and $64$ $3\!\times\!3$ filters per channel, respectively, and again one $1 \times 3$ max-pooling layer. Padding is applied to keep the same dimensions after the convolutions. 
The extracted feature maps are then reshaped into a $N_t \times 3,\!648$ matrices, then fed into a LSTM layer with an output space of dimension $N_t \times 40$. 
Finally, the output layer is a $6$-unit softmax layer that maps each $40$-D vector coming from the LSTM layer into the probability distribution of the number of speakers. After each convolutional layer, rectified linear unit (ReLU) activations are used, whereas in the LSTM cells, we use tanh activation except for the recurrent step which uses hard-sigmoid.

Note that the temporal dimension is preserved all throughout the network to give an output for each frame. In particular for the LSTM layer, as already stated in Section~\ref{seq2seq}, we used sequence-to-sequence decoding so that one output vector corresponds to one input frame of the input tensor. This contrasts with \cite{StoterCountNetEstimatingNumber2019} where sequence-to-one decoding was used in the LSTM to find the maximum number of simultaneous speakers within a $5$-second audio mixture. 
Our sequence-to-sequence set-up is driven by our goal to estimate the number of speakers at framewise resolution.

%\vspace*{-0.2cm}
\section{Experiments}
\label{sec:experiment}

\subsection{Data}
To train and test our CRNN network, we generated a dataset of synthesized speech mixtures with 0 (only spatially diffuse noise) up to $5$ speakers, who speak at random times all along the signal. Inspired by the general methodology of \cite{PerotinCRNNbasedjointazimuth2018}, we used the spatial room impulse response (SRIR) generator \cite{HabetsRoomimpulseresponse2006} to simulate $10,\!200$ ``shoebox'' room configurations  ($10,\!000$ for training, $100$ for validation and $100$ for test). Room length, width and height were randomly chosen within $[2, 10]$\,m, $[2, 10]$\,m and $[2, 3]$\,m, respectively. The reverberation time $T_{60}$ was randomly set between $200$\,ms and $800$\,ms. To simulate up to $5$ speakers in the same room, for each room we generated SRIRs from $5$ different positions with respect to a spherical microphone array randomly located in the room at least $0.5$\,m from the walls. This yielded a total number of $50,\!000$ SRIRs for training, $500$ for validation, and $500$ for test.

We used $16$-kHz speech signals from the TIMIT dataset \cite{timit}. This dataset contains English-speaking sentences uttered by different speakers with different accents. For each room configuration, we created a set of speech mixtures of $15$\,s length with a maximum number of speakers $N_{sp}$ being between $1$ to $5$. The general principle is to first create one-speaker signals by concatenating short sentences of one speaker with transitional silences, then mix those signals together. More specifically, a mixture signal with at most $N_{sp}$ active speakers is generated as follows: i) randomly select a speaker; ii) initialize the signal with a silence of random length within $[0.5, 1]$\,s; iii) randomly pick a sentence from that speaker, concatenate it with the signal, and add a silence of random length within $[0.5, 2]$\,s; iv) repeat step iii) until a $15$\,s signal is obtained. If too long, the final sentence is cropped and faded out in the last $100$\,ms. Finally, v) convolve this dry single-speaker signal with one of the generated random SRIRs. The same procedure is repeated for the $N_{sp}-1$ remaining speakers and $N_{sp}-1$ other SRIRs from the same room to obtain $N_{sp}$ single-speaker reverberant signals. Finally, the $N_{sp}$ signals are added, plus a diffuse noise, to produce a quite realistic reverberant ``conversational'' signal, with a number of speakers varying between $0$ to $N_{sp}$. 

As a result of the above mixture generation (and due to the intermittent nature of speech), frames with a large number of speakers are less likely to occur than the ones comprising none or few speakers. Therefore, to generate a more class-balanced dataset, the probabilities to generate a mixture (for a given room) where $N_{sp}=1$, $N_{sp}=2$, $N_{sp}=3$, $N_{sp}=4$, $N_{sp}=5$ are respectively set to $0.2$, $0.3$, $0.4$, $0.5$ and $1$.

The signal-to-interference ratio (SIR) between the first source and the other sources is set randomly within $0$--$10$~dB. The diffuse noise sequences used in the mixture are of various kinds (crowd, traffic, engine, nature sounds) from the freesound dataset.\footnote{https://freesound.org/} We use a random signal-to-noise ratio (SNR) between 10 and 20~dB with respect to the first source. The diffuse field was simulated by averaging the diffuse parts of two random SRIRs measured in a real reverberant room.

In addition to the control of the inserted silences, we used the TIMIT word timestamps to detect silences within each sentence at sample resolution, and we used them to create frame-wise resolution labels for our dataset. A frame-level label is defined as the maximum number of speakers among all the samples of the frame.

We took care of using different speakers and noise sequences in the train, validation and test sets so that the network is evaluated on speakers and noises sequences unseen during training. The overall duration of the generated mixture dataset is $25$ hours for training, and $0.42$ hours for validation and test.

\vspace*{-0.1cm}
\subsection{Configurations}

 In order to assess the influence of the temporal context on the speaker counting performance, we conducted experiments with different values for the number of frames in the input feature $\mathbf{X} \in \mathbb{R}^{N_t \times 513 \times 4}$. We tested $N_t=10$ ($320$\,ms), $20$ ($640$\,ms) and $30$ ($\approx$ $1$\,s).

\vspace*{-0.1cm}
\subsection{Training procedure}

The CRNN was trained using the ADAM optimizer \cite{KingmaAdamMethodStochastic2014} (with learning rate $= 10^{-3}$, $\beta_{1} = 0.9$, $\beta_{2} = 0.999$, $\epsilon = 10^{-7}$). Early stopping was applied with a patience of $50$ epochs by monitoring the accuracy on the validation set, and the training never exceed 300 epochs. We used the Keras framework with Nvidia GTX 1080 GPUs. 

\vspace*{-0.1cm}
\subsection{Metrics and baseline}
\label{sec:baseline}
As we are considering source counting as a classification problem, we use the per-class classification accuracy as a first metric. It measures the percentage of frames in the test set that are predicted in the same class as the ground truth. In addition, we also use the mean absolute error (MAE) per class $k$ \cite{StoterCountNetEstimatingNumber2019}:
\begin{equation}
    MAE(k) = \frac{1}{T(k)} \sum_{t=1}^{T(k)} |\hat{k}(t)-k|,
\end{equation}
where $\hat{k}(t)$ is the predicted class for frame $t$ of ground-truth class $k$, and $T(k)$ is the total number of frames of class $k$ (here, the different frames of class $k$ are arbitrarily ``re-indexed'' from $1$ to $T(k)$ for simplicity of presentation, although they are not all consecutive frames in the mixture signals).

To assess the advantage of using multichannel features, we trained and tested the same CRNN with single-channel input features, using only the $W$ channel.

\begin{table*}[h!]
\centering
\begin{tabular}{cc|cc:cc:cc:cc:cc:cc|}
\cline{3-14}
\multicolumn{1}{l}{}                              &                 & \multicolumn{12}{c|}{\textbf{number of sources}}                                                                                                                                                                                                    \\ 
\multicolumn{1}{l}{}                              &                 & \multicolumn{2}{c:}{\textbf{0}} & \multicolumn{2}{c:}{\textbf{1}} & \multicolumn{2}{c:}{\textbf{2}} & \multicolumn{2}{c:}{\textbf{3}} & \multicolumn{2}{c:}{\textbf{4}} & \multicolumn{2}{c|}{\textbf{5}} \\ \cline{1-2}
\multicolumn{1}{|l}{{$\boldsymbol{N_t}$}}    & \textbf{method} & \textbf{acc.}                & \textbf{MAE}              & \textbf{acc.}               & \textbf{MAE}              & \textbf{acc.}               & \textbf{MAE}               & \textbf{acc.}               & \textbf{MAE}               & \textbf{acc.}               & \textbf{MAE}               & \textbf{acc.}                & \textbf{MAE}               \\ \hline
\multicolumn{1}{|c}{\multirow{2}{*}{10}} & single-CRNN             & 95.56                & 0.05       & 86.47               & 0.14              & 59.36     & 0.44      & 48.11      & 0.56      & 51.86               & 0.52               & 40.73       & 0.70      \\
\multicolumn{1}{|c}{}                             & multi-CRNN           & \textbf{96.71}               & \textbf{0.03}       & \textbf{91.10}               & \textbf{0.09}              & \textbf{67.21}      & \textbf{0.34}      & \textbf{51.53}      & \textbf{0.51}      & \textbf{54.04}              & \textbf{0.49}               & \textbf{45.08}       & \textbf{0.52}              \\ \cdashline{1-14}
\multicolumn{1}{|c}{\multirow{2}{*}{20}} & single-CRNN             &  93.95              & 0.06        & 87.94              & 0.12     & 70.72               & 0.31               & 52.99              & 0.49               & 49.47     & 0.54      & 35.89                & 0.78               \\
\multicolumn{1}{|c}{}                             & multi-CRNN           & \textbf{96.31}               & \textbf{0.04}       & \textbf{90.31}              & \textbf{0.10}     & \textbf{72.89}               & \textbf{0.28}               & \textbf{55.39}              & \textbf{0.46}               & \textbf{51.54}      & \textbf{0.50}      & \textbf{50.84}               & \textbf{0.54}       \\ \cdashline{1-14}
\multicolumn{1}{|c}{\multirow{3}{*}{30}} & single-CRNN               & \textbf{95.79}                & \textbf{0.04}       & \textbf{90.44}               &  \textbf{0.10}             & 67.80     & 0.34      & 53.24      & 0.49      & 53.89               & 0.48               & 36.92       & 0.72      \\
\multicolumn{1}{|c}{}                             & multi-CRNN           & 95.26                & 0.05       & 89.26               & 0.11              & \textbf{71.76}      & \textbf{0.30}      & \textbf{56.37}      & \textbf{0.46}      & \textbf{55.79}            & \textbf{0.46}               & \textbf{51.43}       & \textbf{0.43}      \\ 

\hline

\end{tabular}
\vspace{0.1cm}
\caption{Classification accuracy and MAE for 0- to 5-speaker mixtures, for the different tested values of $N_t$, and for the single-channel and multichannel (FOA) configurations (best results are in bold).}
\label{tab:results}
\end{table*}

\vspace*{-0.1cm}
\subsection{Results}
\label{sec:evaluation}

Table \ref{tab:results} reports the classification accuracy and the MAE obtained for the single-channel and multichannel CRNNs, for $N_t \in \{10, 20, 30\}$. We have trained each neural network 10 times, then we evaluate each of them on the test set and average the obtained results.

Our CRNN seems rather effective for the $0$-and $1$-source mixtures, then its performance gradually decreases with the increasing number of concurrent speakers. The accuracy still remains at a satisfactory level for the multichannel configuration when $N_t \geq 30$, which obtains a score above $51$\% across all classes. 

We see that the use of the multichannel FOA input yields better performance than the single-channel representation. For example, for $N_t=10,20$, the accuracy is always better from multichannel compared to single-channel, and for $N_t=30$ we can see than multichannel leads to either a close performance to single-channel (for $0$ and $1$ speaker) or a significant improvement (for $2$, $3$, $4$, $5$ speakers). The improvement is even better for a large number of sources, so spatial information seems to help the CRNN for better speaker distinction.

Temporal context seems to have a slight impact on the performance. We see that the results for $N_t=20,30$ are slightly better than the results for $N_t=10$, except for $0$ and $1$ speaker for which the performance is almost the same. 

\section{Conclusion}
\label{sec:conclusion}

The proposed multichannel CRNN yields competitive speaker counting performance at a framewise precision which can be very useful for online speech analysis task such as speaker localization or diarization. It performs with high accuracy for $0$-to-$2$-source mixtures, and decent accuracy in the difficult configuration of more than $2$ sources. The presented results indicate that supplying the network with multichannel features leads to a noticeable improvement in counting performance over the corresponding single-channel model. Yet, we conjecture that there is a considerable margin for improvement, \emph{e.g.} by introducing the higher order Ambisonics (HOA) features, or by using a more adequate network architecture.

\vfill\pagebreak
\bibliographystyle{IEEEtran}
\bibliography{refs}

\end{document}